\documentclass[aps,showpacs,floatfix,twocolumn]{revtex4}
\usepackage{graphicx}
\makeatletter
\newbox\slashbox \setbox\slashbox=\hbox{\large$/$}
\def\pslash#1{\setbox\@tempboxa=\hbox{$#1$}
\@tempdima=0.5\wd\slashbox \advance\@tempdima 0.5\wd\@tempboxa
\copy\slashbox \kern-\@tempdima \box\@tempboxa}

\makeatother
\newcommand{\mat}{\left ( \begin{array}{cc}}
\newcommand{\emat}{\end{array} \right )}
\newcommand{\be}{\begin{eqnarray}}
\newcommand{\ee}{\end{eqnarray}}
\newcommand{\ba}{\begin{array}}
\newcommand{\ea}{\end{array}}

\newcommand{\ben}{\begin{eqnarray*}}
\newcommand{\een}{\end{eqnarray*}}
\def\beq{\begin{equation}}
\def\eeq{\end{equation}}

\begin{document}
\title{Is it possible to observe experimentally
a metal-insulator transition in ultra cold atoms?}
\author{Antonio M. Garc\'{\i}a-Garc\'{\i}a}
\affiliation{Physics Department, Princeton University, Princeton,
New Jersey 08544, USA}
\affiliation{The Abdus Salam International Centre for Theoretical
Physics, P.O.B. 586, 34100 Trieste, Italy}
\author{Jiao Wang}
\affiliation{Department of Physics} \affiliation{Beijing-Hong
Kong-Singapore Joint Center for Nonlinear and Complex Systems
(Singapore), National University of Singapore,117542 Singapore.}
\begin{abstract}
It has been recently reported \cite{ant9} that kicked rotators with certain
non-analytic potentials avoid dynamical localization and undergo a
metal-insulator transition. We show that typical properties of
this transition are still present as the non-analyticity is
progressively smoothed out provided that the smoothing is less
than a certain limiting value. We have identified a smoothing
dependent time scale such that full dynamical localization is
absent and the quantum momentum distribution develops power-law
tails with anomalous decay exponents as in the case of a conductor
at the metal-insulator transition. We discuss under what
conditions these findings may be verified experimentally by using
ultra cold atoms techniques. 
It is found that ultra-cold atoms can indeed be utilized for the
experimental investigation of the
metal-insulator transition.
\end{abstract}
\maketitle

The study of a quantum particle in a random potential
\cite{anderson} is one of the cornerstones of modern condensed
matter physics. In its simplest form, namely, a free spin-less
particle in a short-range disordered potential with no
interactions at zero temperature, the combination of the one
parameter scaling theory \cite{one}, the supersymmetry method
\cite{efetov} and numerical simulations \cite{schreiber}  has led
to the following picture: In two and lower dimensions destructive
interference caused by backscattering produces exponential
localization of the eigenstates in real space for any amount
of disorder. As a consequence, quantum transport is
suppressed, the spectrum is uncorrelated (Poisson) and the system
becomes an insulator. In more than two dimensions there exists a
metal insulator transition (usually referred to as Anderson
transition (AT))for a critical amount of disorder. By critical
disorder we mean a disorder such that, if increased, all the
eigenstates become exponentially localized. For a disorder
strength below the critical one, the system has a mobility edge at
a certain energy which separates localized from delocalized
states. Its position moves away from the band center as the
disorder is decreased. Delocalized eigenstates, typical of a
metal, are extended through the sample and the level statistics
agree with the random matrix prediction for the appropriate
symmetry. In three and higher dimensions the AT takes place in a
region of strong disorder only accessible to numerical
\cite{schreiber,multi} simulations. Typical features of the AT
include:

1. The spectrum of the Hamiltonian is scale invariant \cite{sko},
 namely, any spectral correlator utilized to describe the spectral
 properties of the disordered Hamiltonian does not depend on the
 system size. The spectral correlations at the
AT, usually referred to as critical statistics \cite{sko,kravtsov97},
 are intermediate between that of a metal and that of an insulator.

2. Anomalous scaling of the eigenfunction moments,
 ${\cal P}_q=\int d^dr |\psi({\bf r})|^{2q} \propto L^{-D_q(q-1)}$
 with respect to the sample size $L$, where $D_q$ is a set of exponents
 describing the AT. Eigenfunctions with such a nontrivial (multi)
 scaling are usually dubbed multifractals \cite{multi}
 (for a review see \cite{cuevas}).

3. Quantum diffusion is anomalous \cite{huck} at the AT.
 In the metallic limit, up to small weak localization corrections,
 the density of probability is Gaussian-like and the dynamics
 is well described by a Brownian motion. However, as disorder
 increases, localization effects become important and
quantum diffusion slows down. The density of
 probability develops power-law tails with a decay exponent depending
 on the spectrum of multifractal dimensions \cite{huck}.

Unfortunately the experimental verification of the AT is a
challenging task. In the context of electronic systems is
extremely hard to disentangle effects caused by short decoherence
times, electron-electron interactions and phonon-electron
interactions from destructive quantum interference and symmetry,
supposed to be the main ingredients driving the AT.

In recent years ultracold atoms in optical lattices \cite{raizen}
has been utilized to model certain solid state physics systems.
Generically, in these experiments a very dilute almost free gas of
atoms (Cs and Rb) is cooled up to temperatures of the order of
tens $\mu K$ and then interacts with an optical lattice. In its
simplest form, the optical lattice consists of two laser beams
prepared in such a way that the resulting interference pattern is
a stationary plane wave in space. The laser frequency is tuned
close to a resonance of the atomic system in order to enhance the
atom-laser coupling but not too close to avoid spontaneous
emission. In this limit the system laser-atom can be considered as
a point particle in a sine potential, namely, the quantum
pendulum. Additionally if the laser is turned on only in a series
of short periodic pulses the resulting system is very well
approximated by the so called quantum kicked rotor (see
\cite{reviz} for a review) extensively studied in the context of
quantum chaos, \be {\cal H} = p^2/2 +
k\cos(q)\sum_{n}\delta(t-Tn). \ee The classical motion of
this system is diffusive in momentum space. For short time scales,
quantum and classical motion agrees. However quantum diffusion is
eventually suppressed due to interference effects and eigenstates
are exponentially localized in momentum space. This
counterintuitive feature, usually referred to as dynamical
localization \cite{dyn}, was fully understood \cite{fishman} after
mapping the kicked rotator problem onto an short range 1D
disordered system where localization is
well established. The first direct experimental realization of the
kicked rotor was reported in Ref. \cite{raizen}. As was expected,
the output of the experiment (the distribution of the atom
momentum and the energy diffusion as a function of time) fully
agrees with the theoretical prediction of dynamical localization
\cite{fishman}. Finally we remark that, after the pioneering work
of Ref.\cite{raizen}, many other aspects of the physics of a
quantum kicked rotor as the effect of noise and dissipation have
also been investigated \cite{otherexp} by using similar
experimental settings.

The above results do not depend on the exact details of the potential
but only on its ability to produce classical chaotic motion. The
situation is different if the potential is not smooth.
Recently \cite{ant9} it has been reported that a kicked
rotor could avoid full dynamical localization if the smooth
sinusoidal optical potential is replaced with a generic potential
with a logarithmic or step like singularity. It was found that, for
these potentials, the kicked particle has striking similarities
with a free particle in a disordered potential at the AT. Thus
level statistics are given by critical statistics, eigenfunctions
are multifractal and quantum diffusion becomes anomalous.

A natural question to ask is whether this non-analytical kicked
rotor can be realized in experiments. If so, this would be an
ideal setting to test the physics of the AT.
Obviously, in experiments the singularity can
only be approximated. For instance, an optical lattice
potential with an approximate step-like singularity can be
 produced \cite{gar} either by a holographic mask \cite{mask} with precision
$\sigma$ or by adding a limited number of Fourier components.
In both cases the potential is smooth on sufficiently small scales $\sim
\sigma$. That means that for momenta $p_d \gg \hbar /\sigma$ and
times $t_d$ sufficiently long, the microscopic smoothness of the 
potential is at work and
standard dynamical localization should be observed.
On the other hand, for momenta
$p_c$ and times $t_c$ sufficiently short, classical and quantum
results should coincide. In between these two scales
typical
properties of the AT are observed.
The aim of this paper is twofold on the one hand we seek to determine
in what window of $\sigma$ the AT is observed. On the other hand we examine
whether this range is
already experimentally accessible by using ultra cold atoms in optical
lattices.

 The organization of the paper is as follows: In the next section we
introduce a kicked rotor with
two different smoothed out versions of a step potential.
Then we evaluate the rate of energy diffusion and the full momentum
distribution. Finally we establish
 the minimum smoothing required to observe the AT and whether an experimental
verification is realistic with the current, state of the art,
ultra cold atom techniques.

\begin{figure}[!ht]
\includegraphics[width=.95\columnwidth,clip]{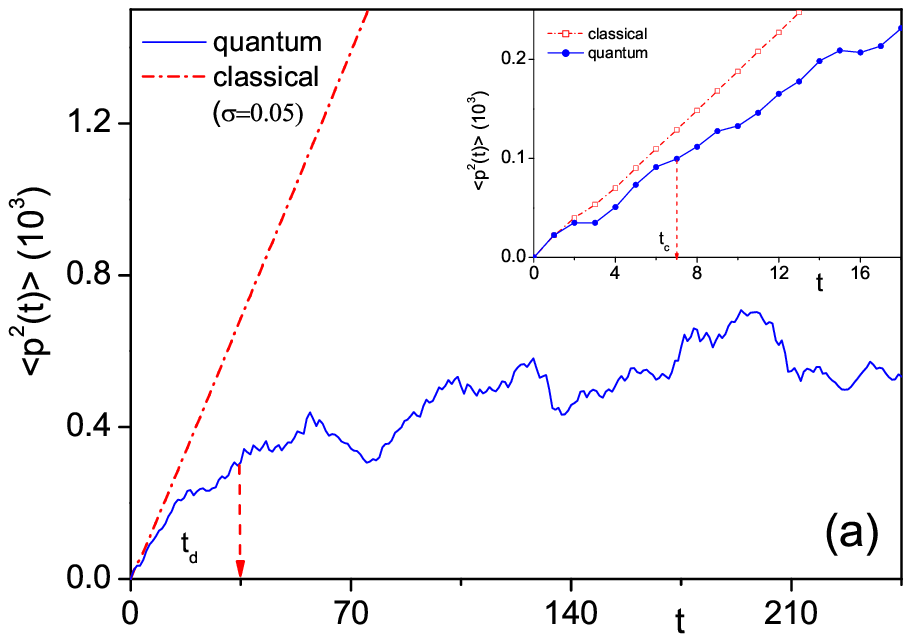}
\vspace{-.5cm}\label{fig1a}
\includegraphics[width=.95\columnwidth,clip]{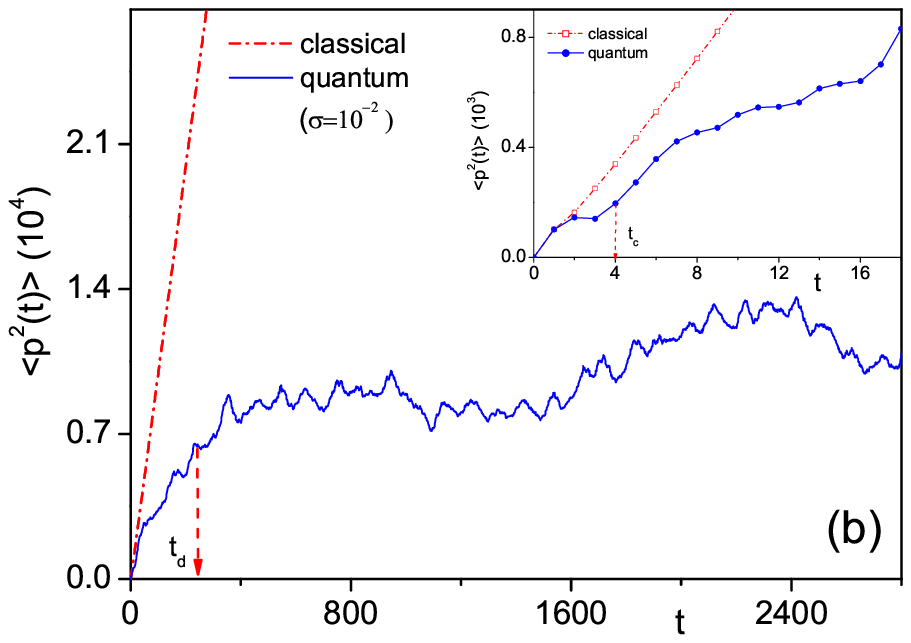}
\vspace{-.5cm}\label{fig1b}
\includegraphics[width=.95\columnwidth,clip]{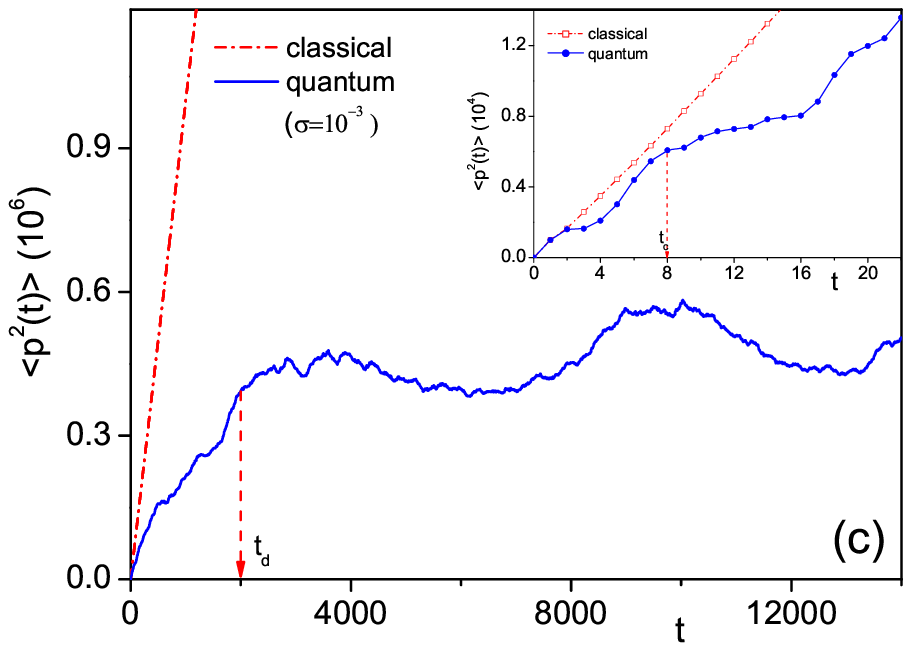}
\vspace{-.5cm}
 \caption{(Color online)Comparison of quantum and
classical energy diffusion versus time for different smoothings:
 $\sigma= 5\times10^{-2}$ (a), $\sigma=10^{-2}$ (b) and
$\sigma=10^{-3}$ (c). Quantum initial conditions, $|\psi(0)\rangle
=|0\rangle$, were chosen to mimic its classical counterpart. For
$t < t_c$  quantum and classical diffusion rates are similar.  For
 $t_d \approx 2/\sigma$ the quantum energy diffusion gets saturated due to
destructive interference. In between these two scales the system
behaves as a disordered conductor at the AT.} \label{fig1}
\end{figure}

\section{The model and observables}
We investigate a  kicked rotor in $1+1$D with a smoothed step-like
potential, \be \label{ourmodel} {\cal H}= \frac{p^2}2
+V_{1,2}(q)\sum_n\delta(t -nT) \ee with $q \in [-\pi,\pi)$. We
consider the following two potentials, \be \label{newmodel} V_1(q)
= Si \left((\frac{\pi}{2}+q)/\sigma \right) + Si
\left((\frac{\pi}{2}-q)/\sigma \right) \ee where $Si(q) =
\int_{0}^{q}\frac{\sin(t)}{t}dt$ is the sine integral function,
and \be \label{newmodel1} V_2(q) = \sum_{m=0}^{M}f(m)\cos(mq) \ee
where $f(m)$ is the discrete Fourier transform of the bare step
like potential $V(q) = \pi$ for $|q| < \pi/2$ and zero otherwise.
In both cases for $\sigma \rightarrow 0$ ($\sigma \equiv 1/M $ in
the latter case) we recover the bare step-like potential
investigated in \cite{ant9}. Obviously there are infinitely many
ways to smooth a singularity, we have chosen the above two due to
similarities with the experimental situation. Thus $V_1(q)$
represents an optical lattice with square-wave intensity profile 
as produced by an array of fine slits or a holographic mask \cite{mask}. The other potential $V_2(q)$
produces an approximated step-like shape by adding a limited
number of Fourier components. We remark that results for $V_1(q)$
and $V_2(q)$ are hardly distinguishable, both are smooth and
oscillatory on scales of the order of $\sigma$. Numerically it is
a little easier to simulate $V_1(q)$ so we will stick to it for
the numerical calculations.

We analyze both the classical and the quantum motion of the
above Hamiltonian. The classical evolution over a period $T$ is dictated by
the map: $p_{n+1}=p_{n}- \frac{\partial V(q_n)}{\partial q_n}$,
$q_{n+1}=q_n+Tp_{n+1}$ (mod$~2\pi$). By smoothing the step
potential the classical force has a well defined classical limit
for any finite $\sigma$.

The quantum dynamics is governed by the quantum evolution operator
$\cal U$ over a period $T$. Thus, after a period $T$, an initial
state $\psi_0$ evolves to $\psi(T) = {\cal U}\psi_0 = e^{\frac{-i
{\hat p}^2T}{2{\bar h}}} e^{-\frac{iV(\hat q)}{\bar h}}\psi_0$
where $\hat p$ and $\hat q$ stands for the usual momentum and
position operator. Our aim is to evolve a given initial state to a
certain time $nT$. This is equivalent to solving the
eigenvalue problem ${\cal U}\Psi_{n}=e^{-i\kappa_n/
\hbar}\Psi_{n}$ where $\Psi_{n}$ is an eigenstate of $\cal U$ with
quasi-eigenvalue $\kappa_n$. In order to proceed we can express
the evolution operator $\langle m| {\cal U} | n \rangle = U_{nm}$
in the basis of momentum eigenstates $\{| n \rangle = \frac {e^{in
\theta}}{\sqrt{2\pi}}\}$ with $n = 0, \ldots N \rightarrow
\infty$, \be \label{uni0} U_{mn} = \frac{e^{-i \frac{T \hbar}{4}
(m^2+n^2)}}{2\pi} \int_{-\pi}^{\pi}dq e^{iq(m-n)-i V(q)/\hbar} \ee
We remark that in this representation, referred to as `cylinder
representation', the resulting matrix  $U_{nm}$ is unitary
exclusively in the $N \rightarrow \infty$ limit. This is certainly
a disadvantage since besides typical finite size effects one has also to
face truncation effects, namely, the integral of the density of
probability is not exactly the unity and eigenvalues are not pure
phases ($e^{-i\theta_n}$) as expected in a Unitary matrix. Moreover
the diagonalization of a generic non Unitary matrix is
numerically much more demanding.

These difficulties can be circumvented by changing representations
in each quantum iteration step, a technique extensively adopted in
quantum kicked rotator studies. First, we express a given state
$\psi$ in position representation, so that it is straightforward
to get $\psi'=e^{-\frac{iV(\hat q)}{\bar h}}\psi$, the state just
after the kick. Next, we express $\psi'$ in the angular momentum
representation by using the fast Fourier transformation (FFT)
algorithm to facilitate the calculation of $ e^{\frac{-i {\hat
p}^2T}{2{\bar h}}} \psi'$.
Since no matrix diagonalization is involved in this scheme, the
computation is quite fast and the effective dimension of the state
vectors is as large as $10^8$. As a result the truncation effects
mentioned above can be safely neglected. We recall that this
method allows to resolve the potential with a precision of
$10^{-8}$, four order less than the minimum $\sigma (10^{-4})$
investigated. Such degree of precision is a necessary requirement
to determine the effect of a small $\sigma$ in the quantum
transport properties of the model studied.

Analytical results for the above model can in principle be
obtained by mapping Eq.(\ref{ourmodel}) onto an 1D Anderson model.
This method was introduced in \cite{fishman} for the case of a
kicked rotor with a smooth potential. We do not repeat here the
details of the calculation but just state how the 1D Anderson
model is modified by the non-analytical potential. It turns out
that the classical non-analyticity induces long-range disorder in
the associated 1D Anderson model. If the kick strength is
sufficiently large the diagonal part of the Anderson model is
pseudo-random and the off-diagonal one decays as $U_r \sim 1/r$
 with $r$ the distance from the diagonal.
This Anderson model is similar to the one studied in \cite{multi}
which is solved by using the supersymmetry method. In general,
according to Ref.\cite{multi}, a $1/r$ decay in 1D is a
signature of an AT. For the potential $V_{1,2}$ above, it is
straightforward to show that $U_r \sim 1/r$ for $r \ll 1/\sigma$
and $U_r \sim e^{-\sigma r}$ for $r > 1/\sigma$. Consequently we
expect to observe AT like behavior for small momentum and then
eventually recover the results of the sinusoidal potential,
namely, exponential localization in momentum space. For further
details of the analytical approach we refer to \cite{ant9}.

We are mainly interested in observables related to transport
properties as the density of probability and the rate of
diffusion.

The density of probability (both classical $P(p,t)$ and quantum
$P_q(p,t)$) of finding a particle with momentum $p$ after a time
$t$ for a given initial state $|\psi(0)\rangle=|0\rangle$.
$P_q(p,t)\equiv P_q(k,t)=|\langle k|\phi(t)\rangle |^2$ with
$p=k\hbar$. In all calculations we set $\hbar=1$. The classical $P(p,t)$ is
obtained by evolving the classical equation of motion for $2\times
10^7$ different initial conditions with zero momentum $p = 0$ and
uniformly distributed position along the interval $(-\pi,\pi)$.
We would like to emphasize our results 
do not depend on the initial conditions. For instance, we have 
checked that similar results are obtained if
 the initial conditions of Ref.\cite{rei2} are considered.

We also examine the second moment of the probability
distribution, namely, the energy diffusion $\langle p^2(t) \rangle
= \int_0^{\infty}dp p^2 P (p,t)$ as a function of time.

We recall our aim is to find out whether the transport properties
are compatibles with those of a disordered conductor at the AT and
how they are affected by the short distance differentiability of
the potential. For the sake of completeness let us briefly summarize
the predictions for both a kicked particle in a smooth potential
and a disordered conductor at the AT.

For a kicked rotator with a smooth potential, it is well
established that initially (up to a certain time $t_c$) both
classical and quantum probabilities are Gaussian like and the
diffusion in momentum is normal, namely, a standard Brownian
motion. For longer times the classical density of probability is
still that of a normal diffusion process. However $P_q(p,t)$
become exponentially localized and energy diffusion stops $\langle
p^2(t) \rangle \sim cons$. These are typical signatures of
dynamical localization.

At the AT, up to a certain $t_c$, agreement is also expected
between the classical and quantum predictions. In the case of a
disordered conductor the classical dynamics is obviously well
described by a Brownian motion. However, for $t > t_c$, the
diffusion becomes anomalous, the quantum density of probability
develop power-law tails in space (localization in a disordered
conductor occurs in real space) and time with exponents related to
the multifractal dimensions of the eigenstates \cite{huck}. The
rate of diffusion is in some cases still similar to the one
corresponding to normal diffusion $\langle p^2(t) \rangle =
D_{quan} t$ though the quantum diffusion constant $D_{quan}$ is
typically lower. This suggests that, at the AT, destructive
interference is still at work but it is not sufficient to fully
localize the particle. In our case we also expect agreement
between classical and quantum results up to a certain time
$t_c(\sigma)$. Additionally,  since the potential is
differentiable for distances smaller than the smoothing $\sigma$,
we expect that there exists a $t_d (\sigma)$ such that for $t \gg
t_d$ standard dynamical localization becomes dominant.

\begin{figure}[!ht]
\includegraphics[width=.95\columnwidth,clip]{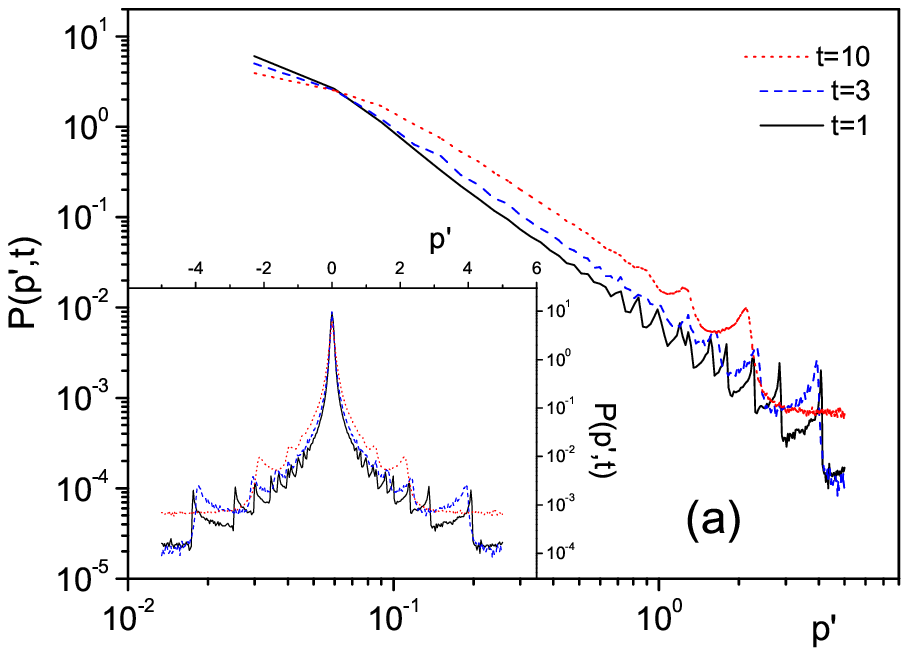}
\vspace{-.5cm} \label{fig2a}
\includegraphics[width=.95\columnwidth,clip]{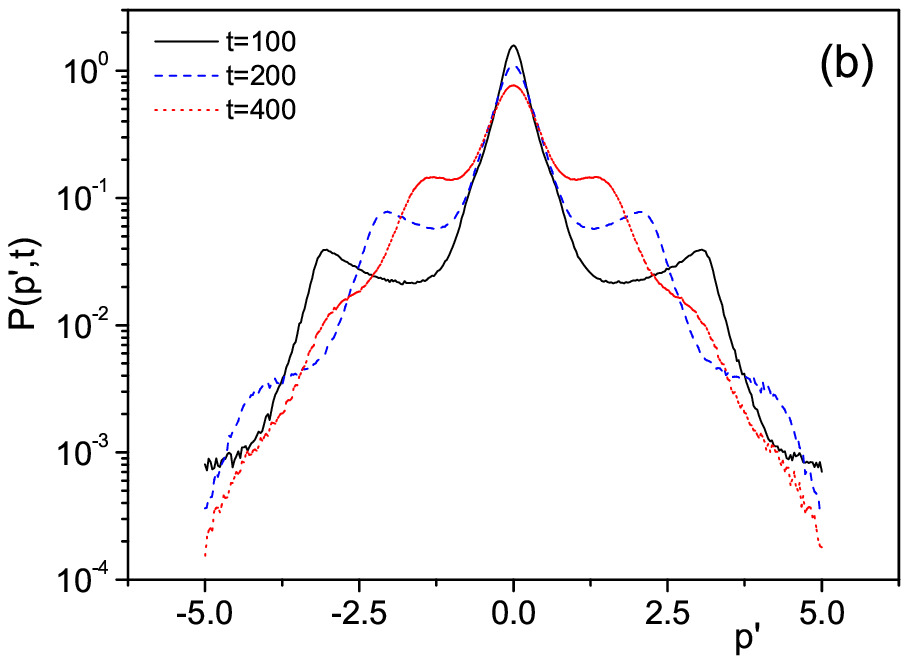}
\vspace{-.5cm} \label{fig2b}
\includegraphics[width=.95\columnwidth,clip]{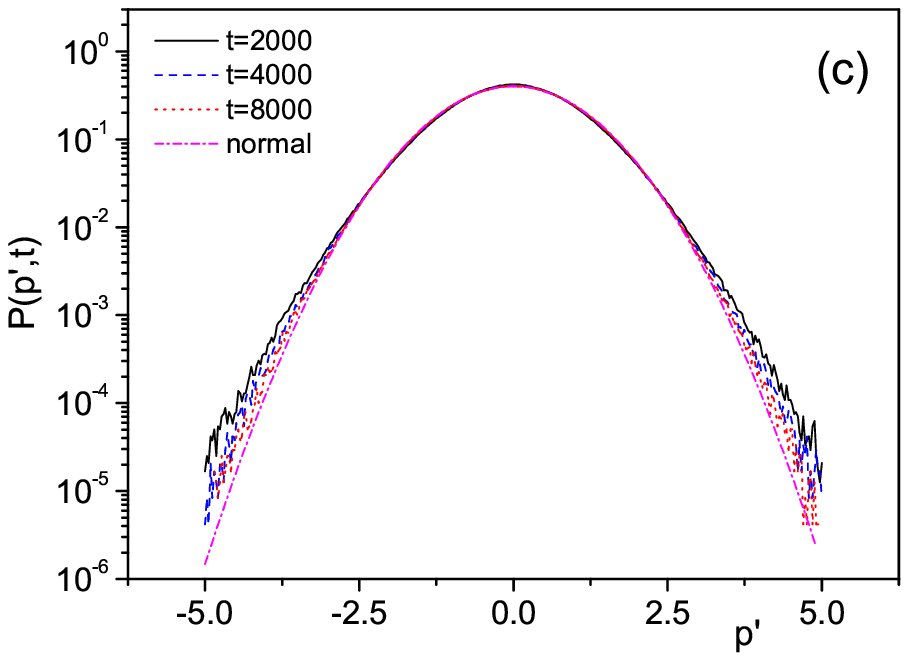}
\vspace{-.5cm} \caption{(Color online) Classical density of
probability distribution $P(p',t)$ for $\sigma=10^{-3}$ and
$p'=p/\sqrt{2Dt}$. (a) Region of anomalous diffusion, $t < 10$,
(b) Crossover from anomalous to normal diffusion $ 10 < t < t_d
\approx 2/\sigma$ and (c) Normal diffusion $t>t_d$. The results
were obtained after averaging over $2\times 10^7$  initial
conditions.} \label{fig2c}
\end{figure}

Typical features of the AT transition are thus observed in our
model only if $t_c \ll t_d$. It is unclear for what range of
$\sigma$, $t_d \gg t_c$ and whether these values of $\sigma$ can
be reached experimentally. We answer these questions in the next
section.

\section{Results}

For the sake of clearness we first enunciate our main conclusions:

1. For $\sigma \leq 0.05$ we have observed typical signatures of an
AT in a broad region of times $t_c \gg t \gg t_d$.

2. The quantum-classical breaking time $t_c$
decreases weakly with $\sigma$. In the range of $\sigma$ investigated
it never goes beyond a few kicks.
 By contrast, the time scale signaling
the beginning of full dynamical localization, due to the
differentiability of the potential, increases as $\sigma$
decreases,  $t_d \approx 2/\sigma$.

3.  We argue that the above range of parameters
is accessible to experimental verification.
By using holographic mask techniques one can reach up to
$\sigma \sim 0.01$ \cite{gar}. On the other hand coherence in ultra cold
atoms  is maintained well beyond $1000$ kicks. Consequently the
AT can be investigated by using ultracold atoms in optical lattices.

\begin{figure}[!ht]
\includegraphics[width=.95\columnwidth,clip]{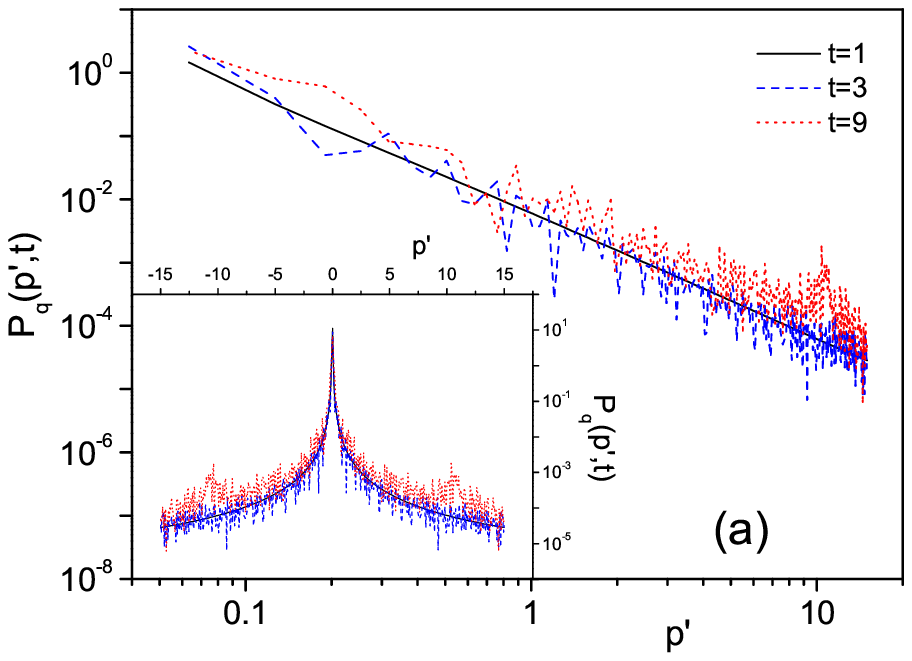}
\vspace{-.5cm} \label{fig3a}
\includegraphics[width=.95\columnwidth,clip]{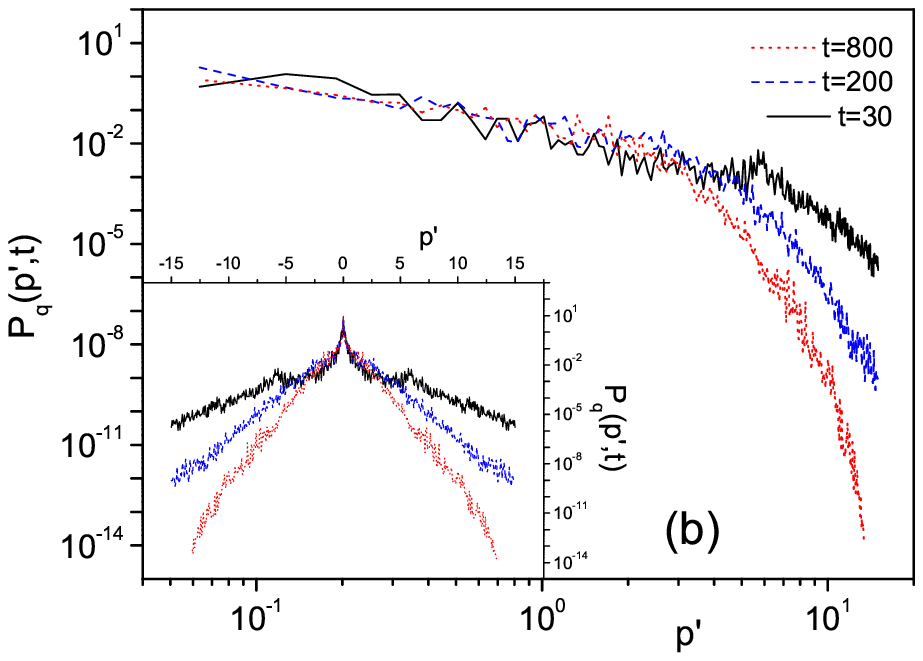}
\vspace{-.5cm} \label{fig3b}
\includegraphics[width=.95\columnwidth,clip]{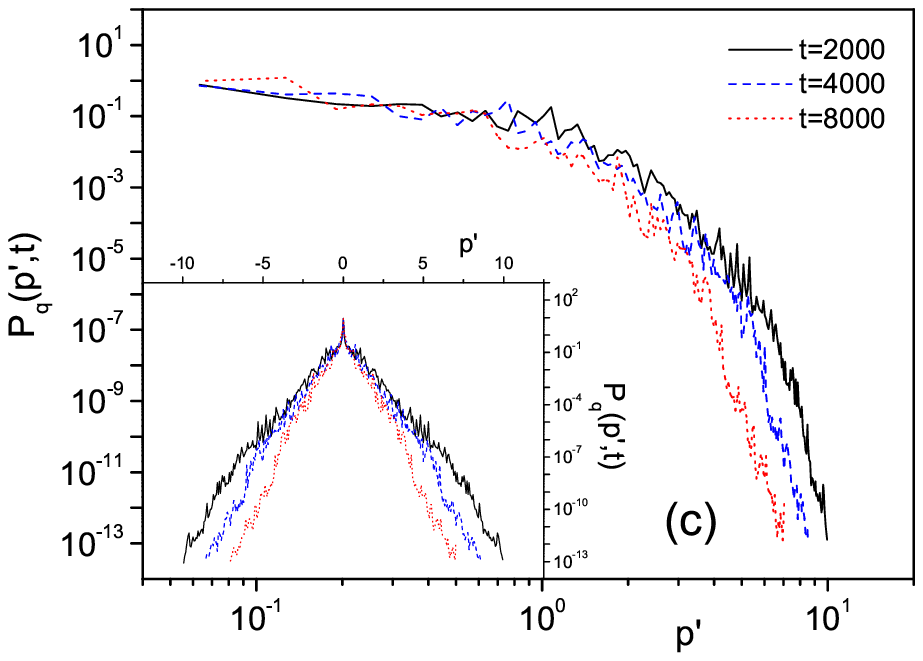}
\vspace{-.5cm} \caption{(Color online) Quantum density of probability
distribution $P_q(p',t)$, $p'=p/\sqrt{2Dt}$ and $\sigma=10^{-3}$
for three different regions: (a) $t < t_c \leq 10$, agreement
between classical and quantum results, (b) $t_c < t < t_d$,
typical properties of an AT are observed, and (c) $t > t_d$,
standard dynamical localization for $|p| > \sqrt{2Dt_d}$. In the
insets we present the same results in a linear-log scale.
$P_q(p',t) \sim 1/p^2$ for $t < t_c$ and $|p|' < 1$, and
$P_q(p',t) \sim 1/p^\alpha$ with $\alpha = 1.1\pm 0.2$ for $t >
t_c$. As initial condition we used $|\psi(0)\rangle
=|0\rangle$.}\label{fig3c}
\end{figure}

We have computed (see details in previous section) the quantum and
classical density of probability for the Hamiltonian  Eq.
\ref{ourmodel} with potentials given  by Eq.\ref{newmodel} and a
variety of smoothings $\sigma \in [0.1,10^{-4}]$.

Our first task is to determine $t_c$ and $t_d$ as a function of
$\sigma$. These time scales can in principle be calculated by
using different observables. Qualitatively all observables should
provide the same physical picture. However the numerical value of
$t_c$ and $t_d$ may depend on the observable considered. For the
sake of simplicity we estimate these time scales by looking at the
the rate of energy diffusion $\langle p^2(t) \rangle$.
\subsection{Energy diffusion}
As was mentioned previously we used as initial conditions $p = 0$
and random positions. Obviously only initial conditions in the
narrow region $[ - \pi + \sigma, \pi - \sigma]$ get a sizable
kick. Thus even after several kicks there is a high probability
that the system stays in the region $p = 0$. In order to show that
our results are not sensitive to initial conditions and stable
under perturbations we have added a weak noise $V(q) = k\sin(q)$
with $k = 1$. We have checked our results do not depend on $k$
provided $k \ll 1/\sigma$. Obviously for $k \sim 1/\sigma$ the
effect of the pseudo-singularity is obscured by the noise
strength.
In the classical case (see Fig.1 )  $\langle p^2(t) \rangle$
increases linearly with time. The dependence of the diffusion
coefficient on $\sigma$ is well approximated by $0.5/\sigma$. This
is consistent with the the analytical prediction resulting from
the random phase approximation \cite{lich}. In the quantum case
(see Fig.1 ) we distinguish three different regions. In a first
stage ($t < t_c \leq 10$) the quantum averaged energy moves around
its classical counterpart. $t_c$ depends weekly on $\sigma$, it
decreases as $\sigma$ does. For longer times $t_c < t < t_d$, the
diffusion is still linear, $\langle p^2(t) \rangle \approx
D_{quan} t$, with $D_{quan} \sim 0.2/\sigma$.
Although it has the same dependence on $\sigma$, it is smaller than in
the classical case. This suggests that quantum interference effects
slow down the classical diffusion. A similar feature has been
found in a disordered conductor at the AT \cite{huck}. This stage lasts up to
$t_d \approx 2/\sigma$. For longer times standard dynamical
localization due to the differentiability of the potential takes
over and diffusion stops.

We recall that linear energy diffusion is only a necessary
condition for normal diffusion. In general the information
obtained from the knowledge of a few moments of the distribution
is not sufficient to fully characterize the classical motion. Thus
the second moment may be $\langle p^2(t) \rangle \sim t$ but this by
no means assures that the density of probability is Gaussian-like
\cite{klafter} as for normal diffusion. We show below that this is
the case in our model.

\subsection{Density of probability}

In the classical case we distinguish two different regimes
separated by a broad crossover region (see density of probability
in Fig.2): First, for short time scales (a few kicks) and
$|p|<c(\sigma)\sqrt{2Dt}$ the diffusion is anomalous. $P(p,t)\sim
p^{-\alpha}$ with $\alpha \sim 2$ and $c(\sigma)\approx 1$
slightly increases as $\sigma$ is decreased. For such a short time
scale the classical system does not feel the differentiability of
the potential.

For longer times but $t < t_d$, we observe a gradual crossover
from anomalous to normal diffusion. For small momentum the density
is still non Gaussian as the effect of the pseudo non
differentiability is still important. As time approaches $t_d$,
the central (small momentum) non-Gaussian region becomes smaller
and smaller. Meanwhile, the outskirts bend down and a
Gaussian-like behavior typical of normal diffusion is observed.
Finally, for $t>t_d$, $P(p,t)$ is well approximated by a Gaussian
distribution. These regions have been observed for all $\sigma$ of
interest.

In the quantum case three regimes are distinguished (see Fig. 3):

1.  $t < t_c$ and $|p| < c(\sigma) \sqrt{2Dt}$ with $D \approx
0.5/\sigma$ ($c(\sigma)$ increases slightly as $\sigma$
decreases). The classical and the quantum probability agree in
this region.
The scale $t_c$ depends weakly on $\sigma$; it
decreases as $\sigma$ does. We recall that our system has not a
well defined classical-quantum correspondence in the limit
$\sigma\rightarrow 0$.
In both cases the diffusion is anomalous $P(p,t)=P_q(p,t) \sim
1/p^2$. We remark $P_q(p,t)$ is calculated by summing the
probability of falling in a bin of width $\Delta p'=0.03$ (so does
for classical $P(p,t)$), hence it is in fact a coarse-grained
result where part of quantum fluctuations have been
suppressed.

2. For $|p| < c(\sigma) \sqrt{2Dt}$ but $t_d > t > t_c$. The
quantum probability $P_q(p,t) \sim 1/p^\alpha$ develop an
power-law tail with an exponent $\alpha < 2$ (see Fig.
3) typical of anomalous diffusion. The
exponent $\alpha$ does not depend on $\sigma$, in all cases we
have found  $ \alpha \sim 1.1\pm 0.2$.
This is a clear signature of an AT. We remark that, in agreement
with previous results from the energy diffusion, the quantum decay
is slower than the classical one. Quantum interference slows down
the motion but it is not enough to fully localize the particle.

3. For $|p| > c(\sigma) \sqrt{2Dt}$ and $t > t_d$, $P_q(p,t)$
decays exponentially. This is an indication of full dynamical
localization due to the differentiability of the potential.

%

\begin{table}

\begin{tabular}{ccc}
  \hline
  $\sigma$ & $t_c$ & $t_d$ \\
  \hline
  $1\times 10^{-1}$ & $~~7 \pm 4$ & $~~~11 \pm 3$\\
  $5\times 10^{-2}$ & $~~7 \pm 4$ & $~~~35 \pm 10$\\
  $2\times 10^{-2}$ & $~~7 \pm 4$ & $~~~50 \pm 25$\\
  $1\times 10^{-2}$ & $~~7 \pm 4$ & $~~~250 \pm 50$\\
  $5\times 10^{-3}$ & $~~8 \pm 4$ & $~~~410 \pm 100$\\
  $2\times 10^{-3}$ & $~~8 \pm 4$ & $~~~1300 \pm 250$\\
  $1\times 10^{-3}$ & $~~8 \pm 4$ & $~~~2000 \pm 400$\\
  $5\times 10^{-4}$ & $~~8 \pm 4$ & $~~~4800 \pm 1200$\\
  \hline
\end{tabular}
\caption{Time scales $t_c$ and $t_d$ for various values of $\sigma$}
\end{table}

From the above we can affirm that in order to observe typical
features of an AT in the transport properties of our system, $t_c$
and $t_d$ must be well separated, namely, $t_d \gg t \gg t_c$. As
is shown in Fig. 3 and Table I, this occurs provided that $\sigma
\leq 0.05$. Thus for an experimental verification of the AT in
cold atoms one has to manage to produce a bare step-potential up
to corrections of order $\sigma \leq 0.05$
\subsection{Experimental verification}
A natural question to ask is what is the minimum values of
$\sigma$ that it can be reached in experiments. Specifically, we
wish to determine, for instance, the maximum number of terms in $V_1(q)$ that can
be included experimentally. In principle \cite{gar} it is an
challenging experimental task to realize optical potentials with
high slopes involving higher optical harmonics of the laser beam.
The problem is that, for instance for Cs, the fourth harmonic is
already in the vacuum UV, and difficult to produce.  Additionally
higher order harmonics are not resonant with the atom and need
a much stronger intensity. Thus it seems extremely hard to go beyond the
first few harmonics. Another option could be to use a kicked rotor
with a smooth potential and three incommensurate frequencies.
According to the results of \cite{shepe}, this model can be mapped
it onto a 3D Anderson model which is supposed to undergo an AT for
a specific value of the coupling constant. However in more than
one dimension there is no clear evidence that this mapping is
really accurate. For instance, the critical exponents at the AT
are very different from the one found in the kicked rotor with
three incommensurate frequencies \cite{she1}.

A more promising alternative is to use a holographic mask to give
a square-wave intensity profile \cite{mask}. This technique combined with the
 recent introduction of spatial light modulators permit the production
of a very broad range of intensity pattern which act as a
effective spatial potential for atoms. Unlike the previous method
the sharpness of the edges would be limited by diffraction effects
of the order of the wavelength \cite{gar}. With the current
techniques the potential of Eq.\ref{newmodel} could be produced in
a window $\sigma \geq 10^{-2}$. On the other hand quantum
coherence in cold atoms is lost after a few thousands kicks.
The experimental bounds are thus within (see above) the theoretical limits
and, as a consequence, the AT can be studied by using ultra cold atoms in
optical lattices.


In conclusion,  we have explicitly shown that kick rotors with a
singular but slightly smoothed potential still have similar
transport properties that those of a disordered conductor at the
AT provided that the degree of smoothing is weak enough.
The utilization of ultra cold atoms in optical
lattices offers the opportunity to investigate Anderson localization in general
and the AT in particular in a setting
free from many of the inconveniences that have plagued other experimental studies of the
AT in the context of condensed matter physics.

\begin{acknowledgments}
AMG thanks D. A. Steck, J. F. Garreau and T. Monteiro for illuminating explanations.
AMG acknowledges financial support from a Marie Curie Outgoing
Fellowship, contract MOIF-CT-2005-007300. JW acknowledges the
support provided by National University of Singapore.
\end{acknowledgments}

\end{document}